\documentclass[
  aps,
  pra,            
  reprint,        
  superscriptaddress,
  amsmath,amssymb,
  longbibliography
]{revtex4-2}

\usepackage{graphicx}      
\usepackage{dcolumn}       
\usepackage{bm}            
\usepackage{braket}
\usepackage{amsmath,amssymb}
\usepackage{bbold}
\usepackage{float}
\usepackage{hyperref}
\hypersetup{    
colorlinks=true,
    linkcolor=blue,
    citecolor=blue,      
    urlcolor=blue
}
\usepackage{orcidlink}
\newcommand{\Hc}{\mathrm{H.c.}}
\newcommand{\half}{\frac{1}{2}}
\newcommand{\tomega}{\tilde{\omega}}
\newcommand{\change}[1]{{\color{black} #1}}

\begin{document}

\title{Collective states of multi-level emitters: The role of multi-level interferences}

\author{Nathan E. Rahat}
\author{Christiane P. Koch\,\orcidlink{0000-0001-6285-5766}}%
\email{christiane.koch@fu-berlin.de}
\affiliation{
Freie Universit\"{a}t Berlin,
Dahlem Center for Complex Quantum Systems and Fachbereich Physik,  Arnimallee 14, 14195 Berlin, Germany}%

\date{\today}

\begin{abstract}
We explore how collective states of light and matter differ when the multi-level nature of the quantum emitters is fully taken into account.
For closely spaced emitters, interferences between near-resonant transitions completely change the character of the collective states compared to two-level approximations. In particular, we find a lower bound on the emitter separation for superradiance to occur which does not exist for two-level emitters. 
By contrast, for larger separations between the emitters, the collective states resemble those obtained within the widely used two-level approximation of the emitters. Both regimes may be realized by molecules trapped in optical lattices. We therefore propose molecular candidates and describe experimental signatures of emerging multi-level interference.
\end{abstract}

\maketitle

\section{Introduction}\label{Sec:intro}

The coupling of atomic or molecular states to the electromagnetic (EM) vacuum is the basis of well-known phenomena such as spontaneous emission. If several atoms or molecules (or more generally, quantum emitters) are arranged in an array whose inter-emitter distances are of the order of an excitation wavelength, they can interact with each other via the EM vacuum to form \textit{collective} states. First introduced by Dicke~\cite{Dicke}, these states have become a paradigmatic example of cooperative behavior, enhancing coherence and giving rise to novel many-body phenomena in light-matter interactions.

Previous studies of collective phenomena have examined the impact of emitter-array geometry on the emergence of collective effects and their spectral properties, such as super- and subradiance~\cite{AsenjoPRX17,Shahmoon17,Holzinger25,Holzinger25_2}. Nowadays, ordered atomic arrays constitute a promising platform not only for studying many-body phenomena~\cite{Masson20}, but also for advancing quantum sensing and metrology~\cite{Hosten16,Colombo22,Shahmoon24} and for the development of new technologies~\cite{Norcia18,Rui2020,Kersten2026}. The theoretical framework used to describe these phenomena relies on a perturbative expansion of the atom-light interaction, resulting in master equations describing the dynamical evolution of the emitter arrays coupled to the EM field~\cite{CohenTannoudjiAtomPhoton}. In these frameworks, the spatial dependence of the EM field is entirely described by the classical Green's tensor, derived from the macroscopic Maxwell's equations~\cite{GrunerWelsch,AsenjoPRA17}. 

A common simplification in theoretical studies of ordered emitter arrays is to treat the emitters as two-level systems. In reality, however, atoms are inherently multi-leveled. \change{While multi-level models have been explored~\cite{AsenjoPRX24,MokGS25,Robicheaux21,Wu2024,Yanes-Thomas25,Konovalov26}, previous studies relied on approximations, such as mean-field treatments, to circumvent the complexity of the associated Hilbert space. When an emitter has two energy levels that are well isolated from the rest, it can be treated as an effective two-level system, justifying the two-level approximation. However, when there are several transitions with closely spaced frequencies, for example when considering the hyperfine structure of an atom or vibronic transitions in a molecule, this approximation breaks down due to the possibility of these transitions to interfere with one another~\cite{Cardimona82,FicekSwain}. For dense atomic systems, this interference changes the resulting spectroscopic signals when the transitions dipoles are parallel~\cite{Horbatsch10,Buchheit16,Konovalov20}. The recent development of experimental platforms capable of realizing emitter arrays with ultra small emitter separations~\cite{holman26} calls for a comprehensive understanding of interference effects in their spectroscopic signals.}

Here we explore how multi-level emitters with close energy levels affect collective states and their dynamical properties, and determine when the two level approximation ultimately fails. We derive a master equation for an array of multi-level quantum emitters (atoms or molecules) interacting with an EM field, allowing different transitions to interfere with one another. The master equation that we obtain resembles those derived from standard volume quantization with the Born–Markov approximation~\cite{Pineiro20} or coarse-graining of time~\cite{Konovalov20}, while additionally enabling a direct study of interference terms between non-parallel transition dipoles without further approximations or assumptions.
With this master equation, we investigate how the multi-level nature of emitters impacts the dynamics of the collective state. Our results indicate that, at very small emitter separations, the multi-level nature of the emitters forming the collective states produces observable deviations from both two-level emitter arrays~\cite{AsenjoNat22,AsenjoPRL20} and arrays with suppressed inter-transition interference~\cite{AsenjoPRX24}. We elucidate the physical conditions under which the multi-level nature of quantum emitters is expected to be significant, and when the two-level approximation will be valid.

The paper is structured as follows: Sec.~\ref{Sec:Theory} presents the mathematical background regarding the master equation describing a multi-level emitter array interacting with an EM field. Sections ~\ref{sec:pop} and \ref{sec:emission} present results of numerical calculations of dynamical properties for a 3 three-level emitter array, exploring the difference between an interfering and non-interfering array. \change{After identifying multi-level effects, Sec.~\ref{sec:validity} discusses the conditions under which they occur, and Sec.~\ref{Sec:examp} presents several possible experimental realizations for observing multi-level interference between near-resonant transitions and collective decay. Section~\ref{Sec:Conclusions} summarizes our findings.}

\section{Theoretical Framework}\label{Sec:Theory}

We consider $N$ multi-level emitters located at positions $\vec{r}_A, ~A\in[1,N]$, interacting with the EM vacuum. The total Hamiltonian consists of three contributions - the emitters' internal structure, the EM field, and the interaction between the two,  
\begin{equation}\label{eq:tot_Ham}
    \hat{H}_{tot}=\hat{H}_{sys}+\hat{H}_{EM}+\hat{H}_{int}\,.
\end{equation}
We adopt the quantization scheme of Refs~\cite{Dung98,Dung02}, which describes the EM field (in atomic units which we use throughout) as
\begin{equation}
    \hat{H}_{EM}=\int d^3r\int_0^\infty d\omega\,\omega\, \hat{\vec{f}}^{\,\dagger}(\vec{r},\omega)\cdot\hat{\vec{f}}(\vec{r},\omega)\,.
\end{equation}
Here, $\hat{\vec{f}}(\vec{r},\omega)$ are bosonic vector operators obeying the standard bosonic 
commutation relations, $[\hat{f}_i(\vec{r},\omega),\hat{f}_j^\dagger(\vec{r'},\omega')]=\delta_{ij}\,\delta(\vec{r}-\vec{r'})\,\delta(\omega-\omega')$, and $[\hat{f}_i(\vec{r},\omega),\hat{f}_j(\vec{r'},\omega')]=[\hat{f}_i^\dagger(\vec{r},\omega),\hat{f}_j^\dagger(\vec{r'},\omega')]=0$.

To describe the emitter's internal structure, we introduce annihilation operators $\hat{\sigma}_{A,k_2\rightarrow k_1}$, describing the de-excitation of emitter $A$ from level $k_2$ to level $k_1$. As a shorthand, a single subscript $k$ refers to a \textit{transition} instead of naming individual levels.
For generality, we make no assumption that the emitters are identical or that their transitions are similar. Hence, the emitters' Hamiltonian is given as 
\begin{equation}\label{eq:sys_ham}
    \hat{H}_{sys}=\sum_{A=1}^{N}\sum_k\half\omega_{A,k}\hat{\sigma}^\dagger_{A,k}\hat{\sigma}_{A,k}\,,
\end{equation}
where $\omega_{A,k}$ is the frequency of the transition $k$ of emitter $A$.

Finally, we describe the light-matter interaction Hamiltonian to lowest order, i.e., neglecting magnetic dipole, electric quadrupole and higher order terms,  
\begin{equation} \label{eq:int_Ham}
\hat{H}_{int}=-\sum_{A=1}^N\sum_k\int_{0}^{\infty}d\omega\left(\hat{d}_{A,k}\cdot\hat{\vec{E}}(\vec{r}_{A},\omega)+\Hc\right)\,,
\end{equation}
where 
\begin{equation}\label{eq:dip_op}
    \hat{d}_{A,k}=\vec{d}_{A,k}\hat{\sigma}_{A,k}+\vec{d}_{A,k}^{\,*}\hat{\sigma}^\dagger_{A,k}\,,
\end{equation} 
with $\vec{d}_{A,k}$ the electric transition dipole moment of the relevant emitter and transition. Unlike in the standard electric-dipole approximation, we are interested in distances smaller than the excitation wavelength and thus need to account for the spatial dependence of the electric field.
Within the EM quantization scheme used here, the spatial dependence of the fields is fully described by the \textit{classical} Green's tensor, derived directly from the macroscopic Maxwell's wave equation~\cite{Dung98}.

\change{Solving the time-dependent Schrödinger equation governed by the total Hamiltonian~\eqref{eq:tot_Ham} is unfeasible. As we are interested only in the dynamics of the emitters, including collective effects, it is sufficient to derive an effective equation of motion for the emitter subsystem while tracing out the EM field's degrees of freedom.
Here, we derive this equation, commonly called a master equation (ME), within the standard Born-Markov approximation to describe the dynamics of the emitters,} 
\begin{align}\label{eq:ME}
    \frac{\partial \hat{\rho}}{\partial t}=-i[\hat{H}_{sys}+\hat{H}_{dd},\hat{\rho}]+\mathcal{L}_{D}(\hat{\rho})\,,
\end{align}
where $\hat\rho(t)$ is the density operator of the emitters at time $t$.
The full derivation of Eq.~\eqref{eq:ME} is provided in appendix~\ref{app:ME}. 

The equation consists of three contributions---the coherent dynamics generated by $\hat H_{sys}$, 
Eq.~\eqref{eq:sys_ham}, the coherent dynamics arising from dipole-dipole (dd) interactions in the emitter array,
\begin{equation}\label{eq:Hdd} \hat{H}_{dd}=-\sum_{A,B=1}^N\sum_{k,m}\Delta_{A,B}^{k,m}\,\hat{\sigma}^\dagger_{A,k}\,\hat{\sigma}_{B,m}\,,
\end{equation}
and the dissipative term,
\begin{eqnarray}\label{eq:Lrho}
  \mathcal{L}_D(\hat{\rho})&=&\half\sum_{A,B=1}^N\sum_{k,m}
  \Gamma_{A,B}^{k,m}
  \bigg(2\hat{\sigma}_{B,m}\hat{\rho}\hat{\sigma}^\dagger_{A,k}\\
  &&\quad\quad-\left\{\hat{\sigma}^\dagger_{A,k}\hat{\sigma}_{B,m},\hat{\rho}\right\} \bigg)\nonumber\,,
\end{eqnarray}
where the curly brackets denote the anti-commutator.
In Eq.~\eqref{eq:Hdd}, $\Delta_{A,B}^{k,m}$ is a scalar prefactor, and for $A=B,k=m$ the Lamb-shift terms are recovered. The contributions with $A=B,k\neq m$ describe interferences between transitions $k,m$ within a single emitter~\cite{Buchheit16}, whereas terms with $A\neq B$ capture dipole-dipole interactions between different emitters and transitions~\cite{Konovalov20,Konovalov26}.
In the dissipator $\mathcal L_D$ in Eq.~\eqref{eq:Lrho}, $\Gamma_{A,B}^{k.m}$ is the rate of the respective decay process. 
Terms with the same emitter and transition index ($A=B$, $k=m$) refer to spontaneous emission from a single level. Terms corresponding to different emitters ($A\neq B$) or transitions ($k \neq m$) refer to correlated decay processes, which form the basis of collective behavior in emitter arrays. 
The coefficients $\Delta_{A,B}^{k,m}$ in Eq.~\eqref{eq:Hdd} and $\Gamma_{A,B}^{k,m}$ in Eq.~\eqref{eq:Lrho} are calculated from
\begin{equation}\label{eq:Coefs_via_GF}   \Delta_{A,B}^{k,m}-\frac{i\Gamma_{A,B}^{k,m}}{2}=-\mu_0\,\omega_{km}^2\,\vec{d}^*_{A,k}\mathbf{G}(\vec{r}_A,\vec{r}_B,\omega_{km})\,\vec{d}_{B,m}\,,
\end{equation}
where $\mathbf{G}(\vec{r}_A,\vec{r}_B,\omega_{km})$ is the classical Green's tensor of the EM field, $\omega_{km}$ is the average of the transition frequencies, $\omega_{km}=(\omega_{A,k}+\omega_{B,m})/2$, and $\mu_0$ is the vacuum's permeability.

The ME~\eqref{eq:ME} reduces to the one obtained for two-level atoms~\cite{AsenjoNat22} when restricting the emitter Hilbert space dimension correspondingly. Previous models of multi-level emitters~\cite{Pineiro20,Pineiro22,Agarwal24,AsenjoPRX24} employed collective raising and lowering operators, i.e., superpositions of individual emitter operators. Instead, we isolate multi-level interference effects by analyzing their role in the dynamics of an emitter array. We therefore retain Eq.~\eqref{eq:ME} without further approximation, enabling a clear separation of interference and non-interference contributions.

In the following, we investigate interference effects through a numerical analysis of the dynamics governed by Eq.~\eqref{eq:ME}. Working with the full equation limits us to small system sizes, while secular error accumulation prevents reliable access to long-time dynamics~\cite{Cattaneo19}. Despite these limitations, multi-level interference leaves clear dynamical signatures. 

\section{Results}\label{Sec:Results}

\begin{figure}
    \centering
    \includegraphics[width=\linewidth]{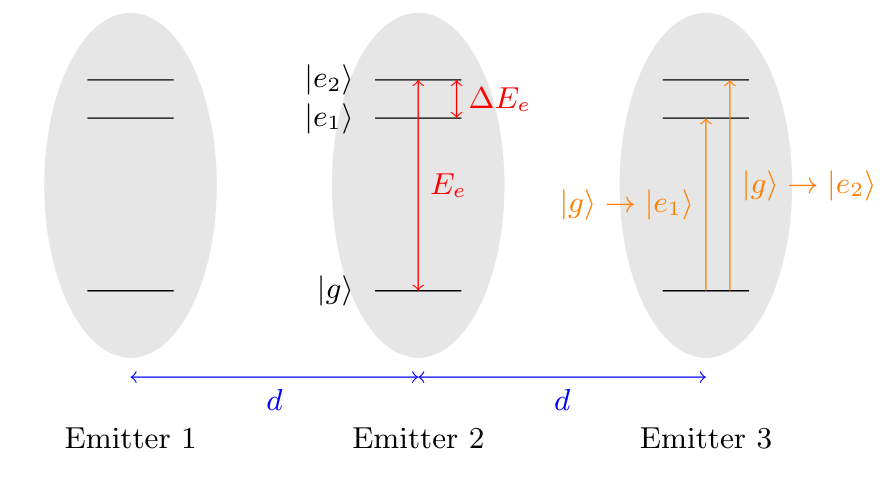}
    \caption{Sketch of a system consisting of three identical emitters arranged in a line with spatial separation $d$ between them, interacting with the quantized EM field. Each emitter consists of three levels in a V configuration - the internal level structure is specified in emitter 2, and the two allowed transitions are indicated in emitter 3.}
    \label{fig:system}
\end{figure}

To study the behavior of an emitter array based on Eq.~\eqref{eq:ME}, we look at the simplest physical scenario that constitutes a multi-level many-emitter array. This allows us to investigate the interference effects that could manifest at a reasonable numerical cost. Namely, we consider an array that consists of three \textit{identical} emitters, whose internal structure consists of three levels. Because the emitters are identical, we can identify the interference contributions of the ME~\eqref{eq:ME} with terms having $k \neq m$, and non-interference terms with $k = m$.

Figure~\ref{fig:system} presents a schematic sketch of the system, consisting of the emitters arranged in a line, separated by distance $d$.
The level scheme of each emitter is that of a V-system, with one ground state $\ket{g}$, and two excited states $\ket{e_1},\ket{e_2}$, with $|E_g-E_{e_1}|<|E_g-E_{e_2}|$. Each emitter has two allowed transitions,  $\ket{g}\leftrightarrow\ket{e_1}$ and $\ket{g}\leftrightarrow\ket{e_2}$, as illustrated on the right emitter in Fig.~\ref{fig:system}. Assuming near-resonant transitions, we define $\Delta E_e$ as the energy separation between the excited states and $E_e$ as the excitation energy from $\ket{g}$ to $\ket{e_2}$, with $\Delta E_e \ll E_e$. The excitation wavelength corresponding to this energy $E_e$ is $\lambda_{e}$. 
The V configuration of a three level emitter corresponds to several physical scenarios. The first one is an electronic excitation in molecular emitters, where an excited electron can end up in one of two vibrational states. Another realization is an atomic excitation into one of two excited hyperfine levels. Here, we start by modeling our emitters based on the $\ket{2s_{1/2},F=0,M_F=0}\rightarrow\ket{4p_{n/2},F=1,M_F=0}$ ($n=1,3$) transitions in hydrogen~\cite{KolachevskyHydrogen18,BeyerHydrogen13}, see appendix \ref{app:Numeric} for further details, but then vary $\Delta E_e/E_e$ to cover further scenarios.

The emitter array interacts with the EM vacuum in free space, such that the Green's tensor is~\cite{GreensF}
\begin{eqnarray}\label{eq:G}
    \mathbf{G}(\vec{r}_{A},\vec{r}_B,\omega)=&&\frac{e^{ikr_{AB}}}{4\pi k^2r_{AB}^3}\bigg[\left(k^2r^2_{AB}+ikr_{AB}-1\right)\mathbb{1}\\\nonumber
&&+    \left(3-k^2r^2_{AB}-3ikr_{AB}\right)\frac{\vec{r}_{AB}\otimes\vec{r}_{AB}}{r_{AB}^2}\bigg]\,,
\end{eqnarray}
where $\vec{r}_{AB}=\vec{r}_A-\vec{r}_B$, $r_{AB}=|\vec{r}_{AB}|$ and $k=\omega/c$.
We use the following state notation: The fully excited state is  $\ket{e_2~e_2~e_2}$, the state with a single excitation on the first emitter is $\ket{e_1~g~g}$ and so on.

In the following, we focus on two scenarios, the full ME~\eqref{eq:ME} and a ME in which interactions between transitions $k \neq m$ are suppressed. The latter corresponds to neglecting any interferences between different transitions.

\subsection{Population dynamics}\label{sec:pop}

\begin{figure*}
  \centering
    \includegraphics[width=0.9\linewidth]{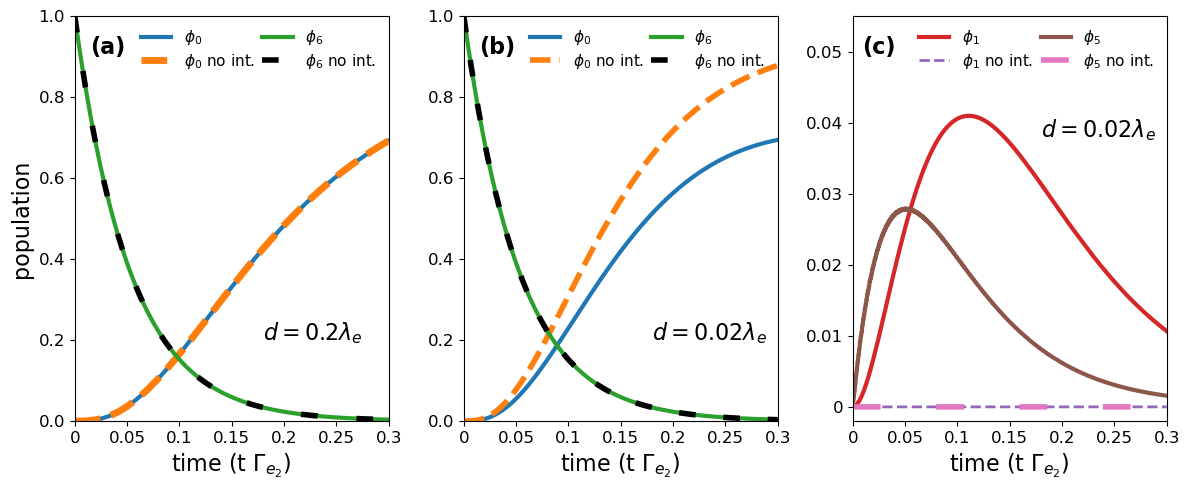}
    \caption{Population dynamics comparing time evolution with (solid lines) and without (dashed lines) interferences between different transitions, the latter obtained by setting all terms with $k\neq m$ to zero in Eq.~\eqref{eq:ME} for two different emitter separations $d$.
    The system is initialized in $\ket{\phi_6}$, and the time units are in terms of the free emitter decay rate from $\ket{e_2}$ to $\ket{g}$.
    }
    \label{fig:pop_dyn}
\end{figure*}

As a first step, we inspect the population dynamics of the emitter array as dictated by Eq.~\eqref{eq:ME}. This elucidates the decay pathways enabled by the interference terms in Eq.~\eqref{eq:ME} and demonstrates their impact on the correlated decay process. We focus on distances shorter than the excitation wavelength. In this regime, the emitters form strongly correlated collective states, such that excitations are delocalized over the array. As a result, the description in terms of individual emitter states is no longer meaningful, and the dynamics are more naturally expressed in a basis of collective states. These correspond to superpositions of individual emitter states with the same energy. We focus in particular on symmetric superpositions, known as Dicke states, as they govern superradiant processes~\cite{GrossHaroche,Nepomechie24}
\begin{eqnarray}\label{eq:Dicke}
    \ket{\phi_0} &=& \ket{g~g~g}\,, \nonumber \\
    \ket{\phi_1} &=& \frac{1}{\sqrt{3}} \big(\ket{e_1~g~g} + \ket{g~e_1~g} + \ket{g~g~e_1} \big)\,, \nonumber \\
    \vdots && \nonumber \\
    \ket{\phi_5} &=& \frac{1}{\sqrt{6}} \big( \ket{e_2~e_1~g} + \ket{e_2~g~e_1} + \ket{e_1~e_2~g} \nonumber \\
             & & \qquad + \ket{e_1~g~e_2} + \ket{g~e_1~e_2} + \ket{g~e_2~e_1} \big)\,, \nonumber \\
    \ket{\phi_6} &=& \ket{e_2~e_2~e_2}\,,
\end{eqnarray}
ordered according to their energy. The array is initialized in the fully excited state, $\ket{\psi(t=0)}=\ket{\phi_6}$. 

The results of the calculation are presented in Fig.~\ref{fig:pop_dyn} as a function of time normalized by the single-emitter decay rate $\Gamma_{e_2}$ of the transition $\ket{e_2}\rightarrow\ket{g}$,.
The population dynamics for the full ME is presented with solid lines, and that of artificially suppressing the multi-level interferences with dashed lines. For a large, yet sub-wavelength, spacing between emitters there is no difference in the predictions when considering inter-level interferences, as shown in Fig.~\ref{fig:pop_dyn}(a) for $d=0.2\lambda_e$. 
When the inter-emitter separation is reduced to $d=0.02\lambda_e$, cf. Fig.~\ref{fig:pop_dyn}(b), both models agree on the population decay of the initial, fully excited state. This is expected since there is no drive to re-excite population into this level. 
Nonetheless, the rate in which the ground state, $\ket{\phi_0}$, is populated shows different behavior between the two scenarios. The full ME dictates a slower population than the non-interfering model. The reason for this difference can be found in Fig.~\ref{fig:pop_dyn}(c), comparing the populations of the states $\ket{\phi_1}$ and $\ket{\phi_5}$ for the interfering and non-interfering scenarios with $d=0.02\lambda_e$. 
While the full ME predicts the population of these states, they are not populated when neglecting multi-level interferences.
This is because both states $\ket{\phi_1}$ and $\ket{\phi_5}$ are composed of individual emitter states that include $\ket{e_1}$ (cf. Eq.~\eqref{eq:Dicke}). Since the system is initialized in a state that involves only the $\ket{e_2}$ internal levels, population cannot be transferred to Dicke states that contain $\ket{e_1}$ without invoking multi-level interferences. The ability to access a larger set of states implies that multi-level interference opens additional decay channels, thereby slowing the general relaxation to the ground state of the array.

\subsection{Emission trends}\label{sec:emission}

\begin{figure}
    \centering
    \includegraphics[width=\linewidth]{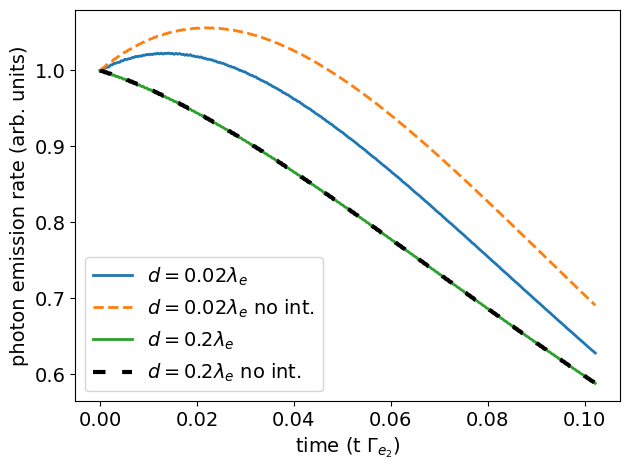}
    \caption{Photon emission rate from the fully excited emitter array. For $d=0.02\lambda_e$, with multi-level interferences (solid blue line) and without (dashed orange line), and $d=0.2\lambda_e$ , with (solid green) and without (dashed brown) interferences.}
    \label{fig:emission}
\end{figure}

The change in population dynamics modifies the total emission from the array. We therefore consider the total photon emission rate of an emitter array initialized in the fully excited state $\ket{\psi(t=0)}=\ket{\phi_6}$. It is given by
\begin{equation}
    S(t)=\sum_{A,B}\sum_{k,m}\Gamma_{A,B}^{k,m}\mathrm{Tr}\bigl\{ \hat{\sigma}_{A,k}^{\dagger}\hat{\rho}(t)\hat{\sigma}_{B,m} \bigr\} \, ,
\end{equation}
where $\Gamma_{A,B}^{k,m}$ are the dissipative coefficients from Eq.~\eqref{eq:Coefs_via_GF}. 

Figure~\ref{fig:emission} displays the total photon emission rate for the two inter-emitter separation values displayed in Fig.~\ref{fig:pop_dyn}. For the array with large inter-emitter separation, $d=0.2\lambda_e$ (solid green and dashed black lines), both the interfering and non-interfering models agree on the emission rate prediction. In contrast, the array with small inter-emitter separation, $d=0.02\lambda_e$ (solid blue and dashed orange lines), presents two different emission rates between the models. Even though both models exhibit superradiance (indicated by the initial increase in emission rate), they disagree on the bursts' strength. The non-interfering model predicts a stronger burst, which is consistent with our finding that it decays more quickly to the ground state.

To quantify the difference in emission between the interfering and non-interfering scenarios, we use the approach proposed in~\cite{AsenjoNat22} to determine if a system exhibits a superradiant burst. Instead of analyzing the entire photon emission pattern, such as the one presented in Fig.~\ref{fig:emission}, we focus on the \textit{initial} photon emission rate, as this is sufficient for determining whether the system exhibits superradiance~\cite{AsenjoNat22}. A superradiant burst occurs only if the first emitted photon increases the probability that the next photons are emitted faster. Hence, if the initial emission rate is positive, the system displays superradiant behavior
\begin{equation}
    \gamma\equiv\frac{dS}{dt}(t=0)>0 \rightarrow \mathrm{superradiance}\,. \nonumber
\end{equation}

\begin{figure}
    \centering
    \includegraphics[width=\linewidth]{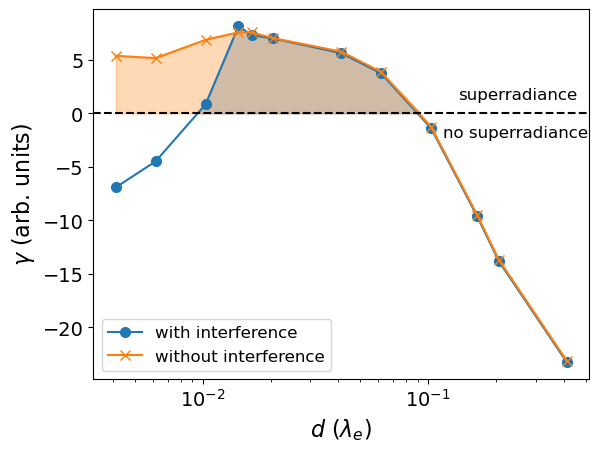}
    \caption{The initial photon emission rate from the array as a function of the spatial distance between emitters, with (blue dots) and without (orange crosses) interferences. The dashed black line at 0 represents the criteria for superradiance.}
    \label{fig:initial_emm_rate}
\end{figure}

The initial emission rate is shown for several values of inter-emitter separations, $d$, in Fig.~\ref{fig:initial_emm_rate} for the interfering and non-interfering cases. The black dashed line corresponds to the superradiance criterion, data above it indicates that the system exhibits a superradiant burst, emphasized also by shading. For inter-emitter separations for which no superradiance occurs, both the interfering and non-interfering models coincide in their emission rates. This trend continues as both predict that a superradiant burst will occur at distances shorter than $0.1\lambda_e$, but when the emitters are very close ($\sim 0.02\lambda_e$) the two scenarios diverge. The interfering scenario predicts that for ultra-small distances (less than $0.01\lambda_e$) the system does not exhibit a superradiant burst, whereas it does for the non-interfering array.
This finding indicates an important feature pertaining to multi-level emitters: When the inter-emitter separation becomes small, superradiance is hindered, and may even disappear completely due to interference between the multiple decay channels.  

\subsection{Range of validity}\label{sec:validity}

\change{Next, we identify more generally the regime in which spectral signatures of multi-level interference become significant, going beyond the hydrogen example discussed thus far.} Two primary parameters determine the emergence of these effects: The spatial distance between the emitters in the array, $d$, and the energy difference between the transitions, $\Delta E$. 
So far, the focus of our investigation has been the spatial separation. Here, we inspect the impact of $\Delta E_e$ on the manifestation of interference effects on the spectral properties of the array. To do so, we inspect again the initial emission rate of the array, but as a function of the energy separations, or "detunings", $\Delta E_e$. We quantify it in terms of the difference between the interfering and non-interfering scenarios
\begin{equation}\label{eq:Delta_S}
    \Delta \gamma\equiv \gamma_{with}-\gamma_{without}\,, \nonumber
\end{equation}
In Fig.~\ref{fig:initial_emm_rate}, this corresponds to the difference between the blue and orange data points.

\begin{figure}
    \centering
    \includegraphics[width=\linewidth]{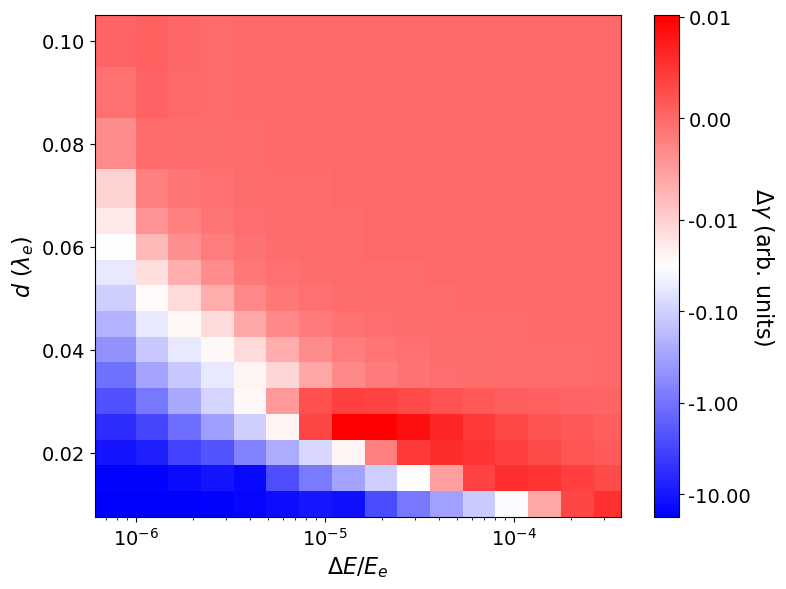}
    \caption{Difference of initial photon emission rate between interfering and non-interfering emitter arrays as a function of the excited state energy separation and the spatial separation within the array.}
    \label{fig:validity_energy}
\end{figure}

The difference between the initial emission rates as a function of both the detunings and emitter separations is presented in Fig.~\ref{fig:validity_energy}. For detunings larger than $3\cdot10^{-4}$ of the main excitation energy, no interference effects are observed. This absence is expected, as interference can only occur between transitions that are energetically similar~\cite{FicekSwain}. As the energy separation decreases, interference effects become more pronounced. The difference between the rates also depends on the inter-emitter separation, $d$. For large $d$, the interfering and non-interfering cases yield similar results regardless of the emitters' internal structure. As the inter-emitter separation gets smaller, the more interference affects the predicted superradiance, and the larger $\Delta E_e$ can be for a difference to be observable. The results presented here indicate that multi-level interferences need to be taken into account for array spacings $0.01\lambda_e\leq d\leq 0.05\lambda_e$, and for emitters whose allowed transitions have an energy separation of $\Delta E_e \leq 10^{-5}E_e$.

\subsection{Experimental candidates}\label{Sec:examp}

\begin{table*}
    \centering
    \begin{tabular}{|c|c|c|c|c|}

        \hline
        System & Transition type & Initial state & Final states & Maximal distance\\
        \hline \hline
        H~\cite{KolachevskyHydrogen18,BeyerHydrogen13}  & electronic & $\ket{2s_{1/2},F=0,M_F=0}$ & \shortstack{$\ket{4p_{1/2},F=1,M_F=0}$ \\ $\ket{4p_{3/2},F=1,M_F=0}$} & 10 nm\\
        \hline
        $^{87}$Rb~\cite{SteckRb01} & electronic & $\ket{5s_{1/2},F=2,M_F=0}$& \shortstack{$\ket{5p_{3/2},F=3,M_F=0}$\\ $\ket{5p_{3/2},F=1,M_F=0}$}& 16 nm\\
        \hline
        $^{15}\mathrm{NH}$~\cite{BailleuxNH12} & rotational & $\ket{N=0,J=1,F=1,F_1=1/2}$&\shortstack{$\ket{N=1,J=1,F=1,F_1=1/2}$\\$\ket{N=1,J=1,F=2,F_1=3/2}$} & 7 $\mu$m\\
        \hline
        IF~\cite{HoeftIF88} & rotational &$\ket{N=0,J=16,F=15.5}$ & \shortstack{$\ket{N=1,J=17,F=16.5}$\\$\ket{N=1,J=17,F=14.5}$}& 22 $\mu$m\\
        \hline

        \end{tabular}
    
    \caption{Examples of atomic and molecular systems which interference effects are expected (the relevant transitions and the maximal distance are noted).}
    \label{tab:examples}
\end{table*}

In the derivation of the ME~\eqref{eq:ME} we have not assumed a specific emitter type, transition type, or energy scale. Hence, it enables prediction of collective effects going beyond the common hydrogen-like atomic candidates including alkaline-earth atoms and molecular emitters. Molecular emitters naturally possess different degrees of freedom across vastly different energy scales which sits within the criteria we presented of multi-level interferences to be significant. 

Table~\ref{tab:examples} presents several candidates for which experimental observations should be possible, together with the relevant transitions within them. Among these are rotational transitions of diatomic molecules, for which we expect multi-level interference to affect the collective decay. These effects are expected to manifest at distances of several microns, a regime accessible with current state-of-the-art optical trapping techniques. The primary experimental challenge in this context is achieving sufficiently low temperatures for trapping the molecular candidates, \change{but current progress in the laser cooling of molecules~\cite{Tarbutt18} suggests that this is not a fundamental obstacle.}

Table~\ref{tab:examples} also includes atomic systems for which multi-level interferences can play a role in high-precision spectroscopy. The systems presented involve electronic transitions in the visible range of the EM spectrum, for which multi-level interference is expected to be significant at separations of only several nanometers.  \change{These distances can be realized experimentally using emitters adsorbed on a surface. The presence of dielectric surroundings is known to affect the coupling between emitters, as the Green's tensor changes based on the surface's dielectric properties~\cite{Olmos18}. Although the presence of a surface adds additional loss channels into the dynamics, collective decay has recently been observed for molecular emitters on a dielectric 2D material~\cite{Juergensen23,Juergensen26}. Since surfaces  naturally support dense packing of the emitters, this platform is a good candidate for  observing spectral signatures of interference.}

\section{Conclusions}\label{Sec:Conclusions}

We have studied collective states in multi-level emitter arrays where near-resonant transitions interfere with each other. The theoretical framework developed here does not require aligned transition dipole moments and is thus suitable not only for atomic, but also for molecular emitters. Our work is a first step towards understanding collective effects in molecular arrays, where the presence of many degrees of freedom and non-parallel transition dipoles is inherent. Since no additional approximations were introduced beyond the standard ones needed to derive the master equation, our results may serve as a benchmark, for example for mean-field treatments or the two-level approximation.

In particular, we have shown that for small distances multi-level interferences change the decay dynamics of the array, enabling new decay paths in the multi-level system. As a result, the total emission from such arrays is diminished in comparison to predictions neglecting the interferences. 
The presence of interference effects is determined by two parameters, the spatial distance between the emitters and the frequency difference between the near-resonant transitions. We have found interference to affect the dynamical properties of an array for spatial distances about two orders of magnitude smaller than the excitation wavelength and for energy differences of around five orders of magnitude smaller than the excitation energy. Experimentally, these requirements can be fulfilled with, e.g., rotational transitions in molecules or electronic transitions within atomic arrays~\cite{holman26} with optical trapping using infrared light. 

Two directions arise for future work exploring interference effects due to the multi-level nature of quantum emitters. The first considers optically trapped emitters in the gas phase, where the influence of array size and dimensionality on the resulting dynamics could be explored using approximations beyond those employed here. Another interesting avenue is the directionality of the emitted light from such arrays~\cite{Carmichael00,AsenjoPRX24}, exploring the impact of the transition dipoles orientation on the resulting emission patterns. Moreover, for molecular arrays in particular, more than two transitions could interfere. The second direction will be to investigate molecular emitters on metallic~\cite{Olmos18} or dielectric surfaces~\cite{Juergensen23,Juergensen26} where the dynamics will be characterized by a competition between light-matter interaction and the various non-radiative decay channels. Combined with the ability to place molecular emitters on nanoscale dielectric structures, 
collective states of light and matter could realize a novel platform for quantum sensing and imaging.

\begin{acknowledgments}
We would like to thank Ronnie Kosloff for fruitful discussions. N.E.R acknowledges support by the IMPRS for Elementary Processes in Physical Chemistry.
\end{acknowledgments}

\appendix

\section{Derivation of the master equation}\label{app:ME}

Our derivation of Eq.~\eqref{eq:ME} utilizes the approach of Ref.~\cite{Dung02_ME}, generalizing it to multi-level emitters.
Starting from the Liouville-von Neumann equation in the Heisenberg picture,  the time evolution of a system operator $\hat{O}$ is generated by the total Hamiltonian, Eq.~\eqref{eq:tot_Ham}.
Dividing the frequency integral in Eq.~\eqref{eq:int_Ham} into resonant (denoted by $\int'd\omega$) and off-resonant (denoted by $\int''d\omega$) contributions, the Liouville-von Neumann equation reads 
\begin{eqnarray}\label{eq:app_O}
  \frac{d}{dt}\hat{O}&=&-i[\hat{O},\hat{H}_{res.}]\\
                     &&+i  \int_0^{''\infty}d\omega\sum_{A,k} \Bigl\{ [\hat{O},\hat{d}_{A,k}]\cdot  \hat{\vec{E}}(\vec{r}_A,\omega)\nonumber\\ 
                    &&\quad\quad\quad\quad+\hat{\vec{E}}^\dagger(\vec{r}_A,\omega)[\hat{O},\hat{d}_{A,k}] \Bigl\}\nonumber\,,
 \end{eqnarray} 
 where \change{we grouped together all contributions from the system and on-resonant interactions into the Hamiltonian}
\begin{eqnarray}\label{eq:H_res}
    \hat{H}_{res.}&=&\int d^3r \int_0^{'\infty}d\omega \,\omega \,\hat{\vec{f}}^{\,\dagger}(\vec{r},\omega)\hat{\vec{f}}(\vec{r},\omega)\\
                &&+\hat{H}_{sys}\nonumber\\
                &&-\int_0^{'\infty}d\omega\sum_{A}\sum_{k}\left(\hat{d}_{A,k}\cdot\hat{\vec{E}}(\vec{r}_A,\omega)+\Hc\right)\,.\nonumber
\end{eqnarray}
$\hat{\vec{E}}(\vec{r}_A,\omega)$ can be expressed in terms of the mode operators $\hat{\vec{f}}(\vec{r},\omega)$~\cite{Dung98},
\begin{equation}
\label{eq:app_E_def} \hat{\vec{E}}(\vec{r},\omega)=2i\,\frac{\omega^2}{c^2}\int d^3r^\prime \sqrt{\varepsilon_I(\vec{r^\prime},\omega)}\mathbf{G}(\vec{r},\vec{r^\prime},\omega)\hat{\vec{f}}(\vec{r^\prime},\omega)
\end{equation}
with $\mathbf{G}(\vec{r}_A,\vec{r},\omega)$ the classical Green's tensor of the EM field and $\varepsilon_I(\vec{r},\omega)$ the imaginary part of the complex (Kramers-Kronig) permittivity. 
The time evolution of the field operators is also dictated by a Liouville von-Neumann equation, 
\begin{eqnarray}\label{eq:f}  
\frac{d}{dt}\hat{\vec{f}}(\vec{r},\omega)&=&-i[\hat{\vec{f}}(\vec{r},\omega),\hat{H}_{EM}+\hat{H}_{int}]\nonumber\\
                                        &=&-i\omega\hat{\vec{f}}(\vec{r},\omega) \\
                                        &&+\frac{2\omega^2}{c^2}\sqrt{\varepsilon_I(\vec{r},\omega)}\cdot\sum_A\sum_k\hat{d}_{A,k}\cdot\mathbf{G}^*(\vec{r}_A,\vec{r},\omega)\,.\nonumber
\end{eqnarray}
Denoting the first term on the r.h.s of Eq.~\eqref{eq:f}, describing the free evolution, by $\hat{\vec{f}}_{free}$ and integrating Eq.~\eqref{eq:f} formally, we obtain 
\begin{eqnarray} 
\hat{\vec{f}}(\vec{r},\omega,t)=\hat{\vec{f}}_{free}(\vec{r},\omega,t)+\frac{2\omega^2}{c^2}\sqrt{\varepsilon_I(\vec{r},\omega)}\cdot\nonumber\\
\sum_A\sum_k\int_0^t dt' \hat{d}_{A,k}(t')\cdot\mathbf{G}^*(\vec{r}_A,\vec{r},\omega)e^{-i\omega(t-t')}\,,\nonumber
\end{eqnarray}
which yields for the electric field
\begin{eqnarray}\label{eq:app_E} 
\hat{\vec{E}}(\vec{r},\omega,t)&=&\hat{\vec{E}}_{free}(\vec{r},\omega,t)
                            \\ &&+4i\,\frac{\omega^2}{c^2}\sum_A\sum_k\int_0^tdt'e^{-i\omega(t-t')}\nonumber\\
                                &&\quad \quad\mathfrak{Im} \mathbf{G}(\vec{r},\vec{r}_A,\omega)\cdot\hat{d}_{A,k}(t')\,,\nonumber
\end{eqnarray}
where we have used Eq.~\eqref{eq:app_E_def} and
$$\frac{\omega^2}{c^2}\int d^3s\varepsilon_I(\vec{s},\omega)\mathbf{G}(\vec{r},\vec{s},\omega)\mathbf{G}^*(\vec{r}',\vec{s},\omega)=\mathfrak{Im}\left( \mathbf{G}(\vec{r},\vec{r}',\omega)\right)\,.$$

Moving into the rotating frame with respect to the emitter energies,  $\tilde{\sigma}_{A,k}(t)=\hat{\sigma}_{A,k}(t)e^{i\omega_{A}t}$, and inserting Eq.~\eqref{eq:app_E} into the second term on the r.h.s of Eq.~\eqref{eq:app_O}, we obtain
\begin{equation}\label{eq:F2}
    F^2(t)\equiv F^2_{free}(t)+\sum_{A,B}F^2_{A,B}\,,
\end{equation}
where $F^2_{free}(t)$ results from $\hat{\vec{E}}_{free}(t)$ and
\begin{widetext}

\begin{eqnarray}
    F^2_{A,B} &=& -4\sum_{k,m} \int_0^t dt' \int_0^{''\infty} d\omega \frac{\omega^2}{c^2}\Big( [\hat{O}(t), \hat{d}_{A,k}(t)] \,\mathfrak{Im}\,\mathbf{G}(\vec{r}_A,\vec{r}_B,\omega) \nonumber \\
    && \times \Big(
        \vec{d}_{B,m}\tilde{\sigma}_{B,m}(t') e^{-i(\omega-\omega_{B,m})(t-t')} e^{-i\omega_{B,m}t} + \vec{d}^*_{B,m}\tilde{\sigma}^\dagger_{B,m}(t') e^{-i(\omega+\omega_{B,m})(t-t')} e^{i\omega_{B,m}t} 
        \Big) \nonumber \\
    && - \mathfrak{Im}\,\mathbf{G}(\vec{r}_B,\vec{r}_A,\omega)
    \Big(
        \vec{d}_{B,m}\tilde{\sigma}_{B,m}(t') e^{i(\omega+\omega_{B,m})(t-t')} e^{-i\omega_{B,m}t}
        + \vec{d}^*_{B,m}\tilde{\sigma}^\dagger_{B,m}(t') e^{-i(\omega-\omega_{B,m})(t-t')} e^{i\omega_{B,m}t}
    \Big) \nonumber \\
    && \times [\hat{O}(t), \hat{d}_{A,k}(t)]
    \Big) \, .
\end{eqnarray}
\end{widetext}

Next, we perform the Markov approximation, $\tilde{\sigma}(t')\rightarrow\tilde{\sigma}(t)$, which is valid on timescales larger than $1/(\omega-\omega_{B,m})$. It allows us to approximate the time integrals as
\begin{eqnarray}\label{eq:xi}
  &&\int_{0}^{t}dt'e^{-i(\omega-\omega_{B,m})(t-t')}\rightarrow \\
  &&\xi(\omega_{B,m}-\omega)\equiv\pi\delta(\omega_{B,m}-\omega)
  +i\mathcal{P}\int_0^\infty\frac{d\omega}{\omega_{B,m}-\omega}\,,\nonumber
\end{eqnarray}
where $\mathcal{P}$ denotes the principal value.
We now apply the secular approximation, i.e., neglect double (de)excitation terms. This approximation must be treated with caution since, in the presence of near-resonant transitions such as those considered here, the assumptions underlying the secular approximation may not be justified~\cite{Cattaneo19,Vaaranta26}. Keeping this caveat in mind, we use the approximation which leads to

\begin{widetext}    
\begin{eqnarray}\label{eq:app_F2}
    F^2_{A,B} &=& -4 \sum_{k,m} \int_0^{''\infty} d\omega \frac{\omega^2}{c^2}
    \Big\{
        \vec{d}_{A,k}\,\mathfrak{Im}\,\mathbf{G}\,\vec{d}^*_{B,m}\,\xi(-(\omega+\omega_{B,m})) [\hat{O},\hat{\sigma}_{A,k}] \hat{\sigma}^\dagger_{B,m} \nonumber\\
        && + \vec{d}^*_{A,k}\,\mathfrak{Im}\,\mathbf{G}\,\vec{d}_{B,m}\,\xi(\omega_{B,m}-\omega) [\hat{O},\hat{\sigma}^\dagger_{A,k}] \hat{\sigma}_{B,m} - \vec{d}^*_{A,k}\,\mathfrak{Im}\,\mathbf{G}\,\vec{d}_{B,m}\,\xi(\omega+\omega_{B,m}) \hat{\sigma}_{B,m} [\hat{O},\hat{\sigma}^\dagger_{A,k}] \nonumber\\
        && - \vec{d}_{A,k}\,\mathfrak{Im}\,\mathbf{G}\,\vec{d}^*_{B,m}\,\xi(\omega-\omega_{B,m}) \hat{\sigma}^\dagger_{B,m} [\hat{O},\hat{\sigma}_{A,k}]
    \Big\}\,.
\end{eqnarray}
where we have omitted the arguments of $\mathfrak{Im}\,\mathbf{G}(\vec{r}_A,\vec{r}_B,\omega)$ for simplicity. 
Since the integration is performed over the off-resonant frequencies, $\omega\pm\tomega_{B,m}\neq0$, and the contribution of the $\delta$-function to $\xi$ can be neglected. Denoting the principal value term by 
\begin{equation}\label{eq:app_Dleta}
    \Delta_{A,B}^{k,m;\pm}=4\, \mathcal{P}\int_{0}^{\infty}d\omega\frac{\omega^2}{c^2}\frac{\vec{d}_{A,k}\mathfrak{Im}\,\mathbf{G}(\vec{r}_A,\vec{r}_B,\omega)\vec{d}_{B,m}}{\omega\pm\omega_{B,m}}\,,
\end{equation}
and inserting Eqs.~\eqref{eq:app_F2} and \eqref{eq:F2} into Eq.~\eqref{eq:app_O}, we obtain 
\begin{eqnarray}\label{eq:app_O2}
  \frac{d}{dt}\hat{O}&=&-i[\hat{O},\hat{H}_{res.}]+F^2_{free}
                    \\&&+i\sum_{A, B} \sum_{k,m} \left\{ 
                        \Delta^{k,m;-}_{A*,B}\left[\hat{O},\hat{\sigma}^\dagger_{A,k}\right]\hat{\sigma}_{B,m}
                        +\Delta^{k,m;+}_{A,B*}\left[\hat{O},\hat{\sigma}_{A,k}\right]\hat{\sigma}^\dagger_{B,m}
                        +\Delta^{k,m;-}_{A,B*}\hat{\sigma}^\dagger_{B,m}\left[\hat{O},\hat{\sigma}_{A,k}\right]
                        +\Delta^{k,m;+}_{A*,B}\hat{\sigma}_{B,m}\left[\hat{O},\hat{\sigma}_{A,k}^\dagger\right] \nonumber
    \right\}\,,
\end{eqnarray}
\end{widetext}
where the notation $A*$($B*$) indicates that $\vec{d}_A$ ($\vec{d}_B$) in Eq.~\eqref{eq:app_Dleta} is replaced by its complex conjugate $\vec{d}_A^*$ ($\vec{d}_B^*$). 
Applying the Kramers-Kronig relation to Eq.~\eqref{eq:app_Dleta}, we find
\begin{displaymath}
\Delta_{A,B}^{k,m;-}=\frac{4\omega_{B,m}^2}{c^2}\vec{d}_{A,k}
  \mathfrak{Re}\mathbf{G}(\vec{r}_A,\vec{r}_B,\omega_{B,m})\vec{d}_{B,m}-\Delta_{A,B}^{k,m;+}\,.  
\end{displaymath}
Next, we take the expectation value of Eq.~\eqref{eq:app_O2} and use the cyclic properties of the trace to arrive at
\begin{widetext}
\begin{eqnarray*}
  \frac{d}{dt}\hat{\rho}&=&
                        -i\left[\hat{H}_{res.},\hat{\rho}\right]
                        +i\sum_{A,B}\sum_{k,m} \left\{ \Delta^{k,m;-}_{A*,B} \left[\hat{\sigma}_{A,k}^\dagger\hat{\sigma}_{B,m}\hat{\rho}-\hat{\sigma}_{B,m}\hat{\rho}\hat{\sigma}^\dagger_{A,k}\right]+\Delta^{k,m;+}_{A,B*}\left[\hat{\sigma}_{A,k}\hat{\sigma}^\dagger_{B,m}\hat{\rho}-\hat{\sigma}^\dagger_{B,m}\hat{\rho}\hat{\sigma}_{A,k}\right] +\Hc\right\}  
\end{eqnarray*}  
\end{widetext}

To simplify this further, we turn to the on-resonance frequency integrals contained in the Hamiltonian $\hat{H}_{res.}$ (Eq.~\eqref{eq:H_res}). These terms can be evaluated using the same steps as above, working in the rotating frame, and employing the Markov and secular approximations.
Since the frequency integrals now run over the resonance region, in approximating $\xi(x)$ we are left with the delta function contribution rather than the principal value integral (cf. Eq.~\eqref{eq:xi}). Denoting
\begin{equation} \Gamma_{A,B}^{k,m}=\frac{8\omega^2_{B,m}}{c^2}\,\vec{d}_{A,k}
  \mathfrak{Im}\mathbf{G}(\vec{r}_A,\vec{r}_B,\omega_{B,m})\,\vec{d}_{B,m}\,,\nonumber
\end{equation}
we arrive at
\begin{widetext}
  \begin{eqnarray}    
  \frac{d}{dt}\hat{\rho}&=&-\frac{i}{2}\sum_A\sum_k\omega_{A,k}
                            \left[\hat{\sigma}^\dagger_{A,k} \hat{\sigma}_{A,k},\hat{\rho}\right]
                            -\half\sum_{A,B}\sum_{k,m}\left\{\Gamma_{A*,B}^{k,m}\left[\hat{\sigma}_{A,k}^\dagger\,\hat{\sigma}_{B,m}\,\hat{\rho}-\hat{\sigma}_{B,m}\,\hat{\rho}\,\hat{\sigma}^\dagger_{A,k}\right]+\Hc\right\} \nonumber\\
                        &&+i\sum_{A, B}\sum_{k,m} \left\{ \Delta_{A*,B}^{k,m;-} \left[\hat{\sigma}_{A,k}^\dagger\,\hat{\sigma}_{B,m}\,\hat{\rho}-\hat{\sigma}_{B,m}\,\hat{\rho}\,\hat{\sigma}^\dagger_{A,k}\right]+\Delta^{k,m;+}_{A,B*}\left[\hat{\sigma}_{A,k}\,\hat{\sigma}^\dagger_{B,m}\,\hat{\rho}-\hat{\sigma}^\dagger_{B,m}\,\hat{\rho}\,\hat{\sigma}_{A,k}\right] +\Hc\right\}\,.\nonumber
\end{eqnarray}
\end{widetext}
To preserve the symmetry of swapping emitters $A$ and $B$, we replace $\omega_{B,m}$ in the coefficients of the ME with $\omega_{km}\equiv(\omega_{A,k}+\omega_{B,m})/2$. This is a reasonable approximation only for near-resonant transitions and only if the Green's tensor varies slowly in the frequency range $|\omega_{A,k}-\omega_{B,m}|$. This replacement results with
\begin{align}
    \Delta_{A,B}^{k*,m;\pm}&\approx\Delta_{B,A}^{m,k*;\pm}\,,\nonumber\\
    \Gamma_{A,B}^{k*,m}&\approx\Gamma_{B,A}^{m,k*}\,.\nonumber
\end{align}
Defining $\Delta_{A*,B}^{k,m}\equiv\Delta_{A*,B}^{k,m;-}+\Delta_{A*,B}^{k,m;+}$ enables us to omit the * superscript, and adopt the convention \change{that, in the notation $\Delta_{A,B}^{k,m}$ and $\Gamma_{A,B}^{k,m}$, the dipole at position $(A,k)$ is always taken to be the conjugated dipole.} Doing so results in Eq.~\eqref{eq:ME} of the main text.

\section{Numerical details}\label{app:Numeric}


The individual emitters constituting our array are modeled as a V-system. In Figs.~\ref{fig:pop_dyn} -~\ref{fig:initial_emm_rate} we use the experimentally measured transition parameters for the hydrogen atom.
Specifically, we consider the $\ket{2s_{1/2},F=0,M_F=0}\rightarrow\ket{4p_{n/2},F=1,M_F=0}$ transitions with $n=1,3$. The associated single emitter radiative shifts are \change{$\Delta_{A}^{e_1}=- 1401.52~\mathrm{kHz}$ and $\Delta_{A}^{e_2}= 1767.30~\mathrm{kHz}$~\cite{KolachevskyHydrogen18,BeyerHydrogen13}. Here we use a shorthand notation for the transitions in which the initial ground state is omitted; for example $\ket{g}\rightarrow\ket{e_2}$ is written simply as $e_2$.} The single emitter cross shift term is calculated to be $\Delta_{A,A}^{e_1,e_2}=\Delta_{A,A}^{e_2,e_1}=366.2~\mathrm{kHz}$~\cite{Buchheit16,Konovalov20}. The corresponding single emitter dissipative coefficients are $\Gamma_{A}^{e_1}=511~\mathrm{kHz}$, $\Gamma_{A}^{e_2}=1022~\mathrm{kHz}$. Unless otherwise noted, the excited state energy separation is set to be $\Delta E_e=1.367~\mathrm{GHz}$. The transition dipole moments are $d_{e_1}=1.28/3~\mathrm{a.u.}$ and $d_{e_2}=\sqrt{2}\cdot 1.28/3~\mathrm{a.u.}$ (obtained from the appropriate Clebsch–Gordan coefficients and the radial overlap integral $\bra{2s}r\ket{4p}=1.28~\mathrm{a.u.}$).
We assume the emitters to be arranged in a line, with the dipoles parallel to each other, and perpendicular to the line of the array. This arrangement simplifies the vacuum Green's tensor,
\begin{align}
    \Delta_{A,B}^{i,j}=\frac{3}{4}\gamma^{A,B}_{i,j}\left( y_0(k_{ij}r_{AB})-\frac{y_1(k_{ij}r_{AB})}{k_{ij}r_{AB}} \right)\,,  \nonumber \\
    \Gamma_{A,B}^{i,j}=\frac{3}{2}\gamma_{i,j}^{A,B}\left( j_0(k_{ij}r_{AB})-\frac{j_1(k_{ij}r_{AB})}{k_{ij}r_{AB}}\right)\nonumber
\end{align}
where $\gamma_{i,j}^{A,B}=(4\omega_{ij}^3d_{A,i}d_{B,k})/(3c^3)$ is a modified Einstein A coefficient. It reduces to its original form for $A=B,~k=m$. $j_q(x)$ and $y_q(x)$ are spherical Bessel functions of the first and second kind, respectively, of order $q$. The wavenumber is $k_{ij}=\omega_{ij}/c$, and $r_{AB}=|\vec{r}_A-\vec{r}_B|$.

The time evolution was calculated by a polynomial approximation of the time evolution operator, $\exp(-i\mathcal{L}\delta t)$, where $\mathcal{L}(\bullet)=-i[H_{sys.}+H_{dd},\bullet]+\mathcal{L}_D(\bullet)$,  using Newton polynomials~\cite{RonnieReview94}.

\bibliographystyle{apsrev4-2}
\bibliography{references}

@Article{AsenjoNat22,
author={Masson, Stuart J.
and Asenjo-Garcia, Ana},
title={Universality of Dicke superradiance in arrays of quantum emitters},
journal={Nature Communications},
year={2022},
day={27},
volume={13},
number={1},
pages={2285},
doi={10.1038/s41467-022-29805-4},
url={https://doi.org/10.1038/s41467-022-29805-4}
}

@article{AsenjoPRX24,
  title = {Dicke Superradiance in Ordered Arrays of Multilevel Atoms},
  author = {Masson, Stuart J. and Covey, Jacob P. and Will, Sebastian and Asenjo-Garcia, Ana},
  journal = {PRX Quantum},
  volume = {5},
  issue = {1},
  pages = {010344},
  numpages = {19},
  year = {2024},
  month = {Mar},
  publisher = {American Physical Society},
  doi = {10.1103/PRXQuantum.5.010344},
  url = {https://link.aps.org/doi/10.1103/PRXQuantum.5.010344}
}

@article{AsenjoPRX17,
  title = {Exponential Improvement in Photon Storage Fidelities Using Subradiance and ``Selective Radiance'' in Atomic Arrays},
  author = {Asenjo-Garcia, A. and Moreno-Cardoner, M. and Albrecht, A. and Kimble, H. J. and Chang, D. E.},
  journal = {Phys. Rev. X},
  volume = {7},
  issue = {3},
  pages = {031024},
  numpages = {36},
  year = {2017},
  month = {Aug},
  publisher = {American Physical Society},
  doi = {10.1103/PhysRevX.7.031024},
  url = {https://link.aps.org/doi/10.1103/PhysRevX.7.031024}
}

@article{AsenjoPRA17,
  title = {Atom-light interactions in quasi-one-dimensional nanostructures: A Green's-function perspective},
  author = {Asenjo-Garcia, A. and Hood, J. D. and Chang, D. E. and Kimble, H. J.},
  journal = {Phys. Rev. A},
  volume = {95},
  issue = {3},
  pages = {033818},
  numpages = {16},
  year = {2017},
  month = {Mar},
  publisher = {American Physical Society},
  doi = {10.1103/PhysRevA.95.033818},
  url = {https://link.aps.org/doi/10.1103/PhysRevA.95.033818}
}

@article{AsenjoPRL20,
  title = {Many-Body Signatures of Collective Decay in Atomic Chains},
  author = {Masson, Stuart J. and Ferrier-Barbut, Igor and Orozco, Luis A. and Browaeys, Antoine and Asenjo-Garcia, Ana},
  journal = {Phys. Rev. Lett.},
  volume = {125},
  issue = {26},
  pages = {263601},
  numpages = {7},
  year = {2020},
  month = {Dec},
  publisher = {American Physical Society},
  doi = {10.1103/PhysRevLett.125.263601},
  url = {https://link.aps.org/doi/10.1103/PhysRevLett.125.263601}
}

@article{Agarwal24,
  title = {Entanglement Generation in Weakly Driven Arrays of Multilevel Atoms via Dipolar Interactions},
  author = {Agarwal, Sanaa and Pi\~neiro Orioli, A. and Thompson, J. K. and Rey, A. M.},
  journal = {Phys. Rev. Lett.},
  volume = {133},
  issue = {23},
  pages = {233003},
  numpages = {7},
  year = {2024},
  month = {Dec},
  publisher = {American Physical Society},
  doi = {10.1103/PhysRevLett.133.233003},
  url = {https://link.aps.org/doi/10.1103/PhysRevLett.133.233003}
}

@article{BailleuxNH12,
	author = {{Bailleux, S.} and {Martin-Drumel, M. A.} and {Margulès, L.} and {Pirali, O.} and {Wlodarczak, G.} and {Roy, P.} and {Roueff, E.} and {Gerin, M.} and {Faure, A.} and {Hily-Blant, P.}},
	title = {High-resolution terahertz spectroscopy of the 15{NH} radical (\hbox{$\widetilde{\rm X}~ ^{3}\Sigma^-$})},
    url =  {https://doi.org/10.1051/0004-6361/201118129},
	journal = {Astron. Astrophys.},
	year = 2012,
	volume = 538,
    pages = {A135}
}

@article{BeyerHydrogen13,
author = {Beyer, Axel and Alnis, Janis and Khabarova, Ksenia and Matveev, Arthur and Parthey, Christian G. and Yost, Dylan C. and Pohl, Randolf and Udem, Thomas and Hänsch, Theodor W. and Kolachevsky, Nikolai},
title = {Precision spectroscopy of the 2S-4P transition in atomic hydrogen on a cryogenic beam of optically excited 2S atoms},
journal = {Annalen der Physik},
volume = {525},
number = {8-9},

url = {https://onlinelibrary.wiley.com/doi/abs/10.1002/andp.201300075},
year = {2013}
}

@article{Buchheit16,
  title = {Master equation for high-precision spectroscopy},
  author = {Buchheit, Andreas Alexander and Morigi, Giovanna},
  journal = {Phys. Rev. A},
  volume = {94},
  issue = {4},
  pages = {042111},
  numpages = {8},
  year = {2016},
  month = {Oct},
  publisher = {American Physical Society},
  doi = {10.1103/PhysRevA.94.042111},
  url = {https://link.aps.org/doi/10.1103/PhysRevA.94.042111}
}

@article{Cardimona82,
doi = {10.1088/0022-3700/15/1/012},
url = {https://dx.doi.org/10.1088/0022-3700/15/1/012},
year = {1982},
volume = {15},
number = {1},
pages = {55},
author = {Cardimona, D. A. and Raymer, M. G. and Stroud, Jr., C. R.},
title = {Steady-state quantum interference in resonance fluorescence},
journal = {Journal of Physics B: Atomic and Molecular Physics},
}

@article{Carmichael00,
title = {A quantum trajectory unraveling of the superradiance master equation1We dedicate this paper to Marlan Scully on the occasion of his 60th birthday.1},
journal = {Optics Communications},
volume = {179},
number = {1},
pages = {417-427},
year = {2000},
issn = {0030-4018},
doi = {https://doi.org/10.1016/S0030-4018(99)00694-X},
url = {https://www.sciencedirect.com/science/article/pii/S003040189900694X},
author = {H.J. Carmichael and Kisik Kim},
}

@article{Cattaneo19,
doi = {10.1088/1367-2630/ab54ac},
url = {https://doi.org/10.1088/1367-2630/ab54ac},
year = {2019},
month = {nov},
publisher = {IOP Publishing},
volume = {21},
number = {11},
pages = {113045},
author = {Cattaneo, Marco and Giorgi, Gian Luca and Maniscalco, Sabrina and Zambrini, Roberta},
title = {Local versus global master equation with common and separate baths: superiority of the global approach in partial secular approximation},
journal = {New J. Phys.},
}

@book{CohenTannoudjiAtomPhoton,
  author    = {Claude Cohen-Tannoudji and Jacques Dupont-Roc and Gilbert Grynberg},
  title     = {Atom-Photon Interactions: Basic Processes and Applications},
  publisher = {Wiley},
  address   = {New York},
  year      = {1992},
  isbn       = {978-0471625568},
  url       = {https://onlinelibrary.wiley.com/doi/book/10.1002/9783527617197}
}

@Article{Colombo22,
author={Colombo, Simone and Pedrozo-Pe{\~{n}}afiel, Edwin and Adiyatullin, Albert F. and Li, Zeyang and Mendez, Enrique and Shu, Chi and Vuleti{\'{c}}, Vladan},
title={Time-reversal-based quantum metrology with many-body entangled states},
journal={Nature Physics},
year={2022},
month={Aug},
day={01},
volume={18},
number={8},
pages={925-930},
doi={10.1038/s41567-022-01653-5},
url={https://doi.org/10.1038/s41567-022-01653-5}
}

@article{Dicke,
  title = {Coherence in Spontaneous Radiation Processes},
  author = {Dicke, R. H.},
  journal = {Phys. Rev.},
  volume = {93},
  issue = {1},
  pages = {99--110},
  numpages = {0},
  year = {1954},
  month = {Jan},
  publisher = {American Physical Society},
  doi = {10.1103/PhysRev.93.99},
  url = {https://link.aps.org/doi/10.1103/PhysRev.93.99}
}

@article{Dung98,
  title = {Three-dimensional quantization of the electromagnetic field in dispersive and absorbing inhomogeneous dielectrics},
  author = {Dung, Ho Trung and Kn\"oll, Ludwig and Welsch, Dirk-Gunnar},
  journal = {Phys. Rev. A},
  volume = {57},
  issue = {5},
  pages = {3931--3942},
  numpages = {0},
  year = {1998},
  publisher = {American Physical Society},
  doi = {10.1103/PhysRevA.57.3931},
  url = {https://link.aps.org/doi/10.1103/PhysRevA.57.3931}
}

@article{Dung02,
  title = {Intermolecular energy transfer in the presence of dispersing and absorbing media},
  author = {Dung, Ho Trung and Kn\"oll, Ludwig and Welsch, Dirk-Gunnar},
  journal = {Phys. Rev. A},
  volume = {65},
  issue = {4},
  pages = {043813},
  numpages = {13},
  year = {2002},
  month = {Apr},
  publisher = {American Physical Society},
  doi = {10.1103/PhysRevA.65.043813},
  url = {https://link.aps.org/doi/10.1103/PhysRevA.65.043813}
}

@article{Dung02_ME,
  title = {Resonant dipole-dipole interaction in the presence of dispersing and absorbing surroundings},
  author = {Dung, Ho Trung and Kn\"oll, Ludwig and Welsch, Dirk-Gunnar},
  journal = {Phys. Rev. A},
  volume = {66},
  issue = {6},
  pages = {063810},
  numpages = {16},
  year = {2002},
  month = {Dec},
  publisher = {American Physical Society},
  doi = {10.1103/PhysRevA.66.063810},
  url = {https://link.aps.org/doi/10.1103/PhysRevA.66.063810}
}

@book{GreensF,
    place={Cambridge}, 
    edition={2}, 
    title={Principles of nano-optics}, 
    publisher={Cambridge University Press}, 
    author={Novotny, Lukas and Hecht, Bert}, 
    year={2012},
    doi = {10.1017/CBO9780511794193},
    url = {https://doi.org/10.1017/CBO9780511794193}
}

@article{GrossHaroche,
title = {Superradiance: An essay on the theory of collective spontaneous emission},
journal = {Physics Reports},
volume = {93},
number = {5},
pages = {301-396},
year = {1982},
issn = {0370-1573},
doi = {https://doi.org/10.1016/0370-1573(82)90102-8},
url = {https://www.sciencedirect.com/science/article/pii/0370157382901028},
author = {M. Gross and S. Haroche},
}

@article{GrunerWelsch,
  title = {Green-function approach to the radiation-field quantization for homogeneous and inhomogeneous Kramers-Kronig dielectrics},
  author = {Gruner, T. and Welsch, D.-G.},
  journal = {Phys. Rev. A},
  volume = {53},
  issue = {3},
  pages = {1818--1829},
  year = {1996},
  month = {Mar},
  publisher = {American Physical Society},
  doi = {10.1103/PhysRevA.53.1818},
  url = {https://link.aps.org/doi/10.1103/PhysRevA.53.1818}
}

@article{HoeftIF88,
    author = {Hoeft, J. and Nair, K.P.R},
    title = {Hyperfine structure, mm wave rotational spectrum and molecular constatns of the diatomic IF in its electronic ground state \hbox{$\mathrm{X}~ ^{1}\Sigma^+$}},
    journal = {Zeitschrift für Physik D Atoms, Molecules and Clusters},
    volume = {8},
    url={https://doi.org/10.1007/BF01384527},
    year = {1988}
}

@misc{holman26,
      title={A Mid-Infrared Platform Based on Strontium Tweezer Arrays}, 
      author={Aaron Holman and Ximo Sun and Bojeong Seo and Joshua Corn and Zezheng Zhu and Yuan Xu and Jiahao Wu and Nanfang Yu and Dmytro Filin and Marianna Safronova and Sebastian Will},
      year={2026},
      eprint={2606.02560},
      archivePrefix={arXiv},
      primaryClass={physics.atom-ph},
      url={https://arxiv.org/abs/2606.02560}, 
}

@misc{Holzinger25,
      title={Superradiant Peak Emission Rate and Time in Quantum Emitter Arrays}, 
      author={Raphael Holzinger and Susanne F. Yelin},
      year={2025},
      eprint={2504.09985},
      archivePrefix={arXiv},
      primaryClass={quant-ph},
      url={https://arxiv.org/abs/2504.09985}, 
}

@misc{Holzinger25_2,
      title={Scaling of Superradiant Peak Emission in Spatially Extended Emitter Arrays}, 
      author={Raphael Holzinger and Susanne F. Yelin},
      year={2025},
      eprint={2506.12649},
      archivePrefix={arXiv},
      primaryClass={quant-ph},
      url={https://arxiv.org/abs/2506.12649}, 
}

@article{Horbatsch10,
  title = {Shifts from a distant neighboring resonance},
  author = {Horbatsch, M. and Hessels, E. A.},
  journal = {Phys. Rev. A},
  volume = {82},
  issue = {5},
  pages = {052519},
  numpages = {6},
  year = {2010},
  month = {Nov},
  publisher = {American Physical Society},
  doi = {10.1103/PhysRevA.82.052519},
  url = {https://link.aps.org/doi/10.1103/PhysRevA.82.052519}
}

@Article{Hosten16,
author={Hosten, Onur and Engelsen, Nils J. and Krishnakumar, Rajiv and Kasevich, Mark A.},
title={Measurement noise 100 times lower than the quantum-projection limit using entangled atoms},
journal={Nature},
year={2016},
day={01},
volume={529},
number={7587},
pages={505-508},
doi={10.1038/nature16176},
url={https://doi.org/10.1038/nature16176}
}

@book{FicekSwain,
    author = {Ficek, Zbigniew and Swain, Stuart},
    title = {Quantum Interference and Coherence: Theory and Experiments},
    publisher = {Springer-Verlag New York},
    year = {2005},
    doi= {https://doi.org/10.1007/b100106}
}

@article{Juergensen23,
author = {Juergensen, Sabrina and Kessens, Moritz and Berrezueta-Palacios, Charlotte and Severin, Nikolai and Ifland, Sumaya and Rabe, J{\"u}rgen P. and Mueller, Niclas S. and Reich, Stephanie},
title = {Collective States in Molecular Monolayers on 2D Materials},
journal = {ACS Nano},
volume = {17},
number = {17},
pages = {17350-17358},
year = {2023},
doi = {10.1021/acsnano.3c05384},
URL = {https://doi.org/10.1021/acsnano.3c05384},
}

@article{Juergensen26,
author = {Juergensen, Sabrina and Marceau, Jean-Baptiste and Mueller, Chantal and Barros, Eduardo B. and Kusch, Patryk and Setaro, Antonio and Gaufrès, Etienne and Reich, Stephanie},
title = {Collective States of α-Sexithiophene Chains Inside Boron Nitride Nanotubes},
journal = {The Journal of Physical Chemistry Letters},
volume = {16},
number = {9},
pages = {2393-2400},
year = {2025},
doi = {10.1021/acs.jpclett.4c02977},
}

@Article{Kersten2026,
author={Kersten, Wenzel
and de Zordo, Nikolaus
and Diekmann, Oliver
and Redchenko, Elena S.
and Kanagin, Andrew N.
and Angerer, Andreas
and Munro, William J.
and Nemoto, Kae
and Mazets, Igor E.
and Rotter, Stefan
and Pohl, Thomas
and Schmiedmayer, J{\"o}rg},
title={Self-induced superradiant masing},
journal={Nature Physics},
year={2026},
month={Jan},
day={01},
volume={22},
number={1},
pages={158-163},
issn={1745-2481},
doi={10.1038/s41567-025-03123-0},
url={https://doi.org/10.1038/s41567-025-03123-0}
}

@article{KolachevskyHydrogen18,
    author = {Kolachevsky, N. and Beyer, A. and Maisenbacher, L. and Matveev, A. and Pohl, R. and Khabarova, K. and Grinin, A. and Lamour, T. and Yost, D. C. and Haensch, T. W. and Udem, Th.},
    title = {2S-4S spectroscopy in hydrogen atom: The new value for the Rydberg constant and the proton charge radius},
    journal = {AIP Conference Proceedings},
    volume = {1936},
    number = {1},
    pages = {020015},
    year = {2018},
    month = {02},
    issn = {0094-243X},
    doi = {10.1063/1.5025453},
}

@article{Konovalov20,
  title = {Master equation for multilevel interference in a superradiant medium},
  author = {Konovalov, Aleksei and Morigi, Giovanna},
  journal = {Phys. Rev. A},
  volume = {102},
  issue = {1},
  pages = {013724},
  numpages = {12},
  year = {2020},
  publisher = {American Physical Society},
  doi = {10.1103/PhysRevA.102.013724},
  url = {https://link.aps.org/doi/10.1103/PhysRevA.102.013724}
}

@misc{Konovalov26,
      title={Vacuum-induced interference in light scattering by multilevel atomic chains}, 
      author={Aleksei Konovalov and Giovanna Morigi and Nicola Piovella},
      year={2026},
      eprint={2607.21073},
      archivePrefix={arXiv},
      primaryClass={quant-ph},
      url={https://arxiv.org/abs/2607.21073}, 
}

@article{Masson20,
  title = {Many-Body Signatures of Collective Decay in Atomic Chains},
  author = {Masson, Stuart J. and Ferrier-Barbut, Igor and Orozco, Luis A. and Browaeys, Antoine and Asenjo-Garcia, Ana},
  journal = {Phys. Rev. Lett.},
  volume = {125},
  issue = {26},
  pages = {263601},
  numpages = {7},
  year = {2020},
  publisher = {American Physical Society},
  doi = {10.1103/PhysRevLett.125.263601},
  url = {https://link.aps.org/doi/10.1103/PhysRevLett.125.263601}
}

@article{MokGS25,
  title = {Ground-state selection via many-body superradiant decay},
  author = {Mok, Wai-Keong and Masson, Stuart J. and Stamper-Kurn, Dan M. and Zelevinsky, Tanya and Asenjo-Garcia, Ana},
  journal = {Phys. Rev. Res.},
  volume = {7},
  issue = {2},
  pages = {L022015},
  numpages = {7},
  year = {2025},
  month = {Apr},
  publisher = {American Physical Society},
  doi = {10.1103/PhysRevResearch.7.L022015},
  url = {https://link.aps.org/doi/10.1103/PhysRevResearch.7.L022015}
}

@article{Nepomechie24,
author = {Nepomechie, Rafael I. and Ravanini, Francesco and Raveh, David},
title = {Spin-$s$ Dicke States and Their Preparation},
journal = {Advanced Quantum Technologies},
volume = {7},
number = {12},
pages = {2400057},
doi = {https://doi.org/10.1002/qute.202400057},
url = {https://advanced.onlinelibrary.wiley.com/doi/abs/10.1002/qute.202400057},
year = {2024}
}

@article{Norcia18,
  title = {Frequency Measurements of Superradiance from the Strontium Clock Transition},
  author = {Norcia, Matthew A. and Cline, Julia R. K. and Muniz, Juan A. and Robinson, John M. and Hutson, Ross B. and Goban, Akihisa and Marti, G. Edward and Ye, Jun and Thompson, James K.},
  journal = {Phys. Rev. X},
  volume = {8},
  issue = {2},
  pages = {021036},
  numpages = {12},
  year = {2018},
  month = {May},
  publisher = {American Physical Society},
  doi = {10.1103/PhysRevX.8.021036},
  url = {https://link.aps.org/doi/10.1103/PhysRevX.8.021036}
}

@article{Olmos18,
  title = {Modified dipole-dipole interaction and dissipation in an atomic ensemble near surfaces},
  author = {Jones, Ryan and Needham, Jemma A. and Lesanovsky, Igor and Intravaia, Francesco and Olmos, Beatriz},
  journal = {Phys. Rev. A},
  volume = {97},
  issue = {5},
  pages = {053841},
  numpages = {13},
  year = {2018},
  month = {May},
  publisher = {American Physical Society},
  doi = {10.1103/PhysRevA.97.053841},
  url = {https://link.aps.org/doi/10.1103/PhysRevA.97.053841}
}

@article{Pineiro20,
  title = {Subradiance of multilevel fermionic atoms in arrays with filling $n\ensuremath{\ge}2$},
  author = {Pi\~neiro Orioli, A. and Rey, A. M.},
  journal = {Phys. Rev. A},
  volume = {101},
  issue = {4},
  pages = {043816},
  numpages = {22},
  year = {2020},
  month = {Apr},
  publisher = {American Physical Society},
  doi = {10.1103/PhysRevA.101.043816},
  url = {https://link.aps.org/doi/10.1103/PhysRevA.101.043816}
}

@article{Pineiro22,
  title = {Emergent Dark States from Superradiant Dynamics in Multilevel Atoms in a Cavity},
  author = {Pi\~neiro Orioli, A. and Thompson, J. K. and Rey, A. M.},
  journal = {Phys. Rev. X},
  volume = {12},
  issue = {1},
  pages = {011054},
  numpages = {36},
  year = {2022},
  month = {Mar},
  publisher = {American Physical Society},
  doi = {10.1103/PhysRevX.12.011054},
  url = {https://link.aps.org/doi/10.1103/PhysRevX.12.011054}
}

@article{Robicheaux21,
  title = {Theoretical study of early-time superradiance for atom clouds and arrays},
  author = {Robicheaux, F.},
  journal = {Phys. Rev. A},
  volume = {104},
  issue = {6},
  pages = {063706},
  numpages = {10},
  year = {2021},
  month = {Dec},
  publisher = {American Physical Society},
  doi = {10.1103/PhysRevA.104.063706},
  url = {https://link.aps.org/doi/10.1103/PhysRevA.104.063706}
}

@Article{RonnieReview94,
author = 	 {Kosloff, R.},
title = 	 {Propagation methods for molecular dynamics},
journal = 	 {Annu. Rev. Phys. Chem.},
year = 	 {1994},
volume = 	 {45},
pages = 	 {145-178},
url ={https://doi.org/10.1146/annurev.pc.45.100194.001045},
doi ={10.1146/annurev.pc.45.100194.001045},
}

@Article{Rui2020,
author={Rui, Jun
and Wei, David
and Rubio-Abadal, Antonio
and Hollerith, Simon
and Zeiher, Johannes
and Stamper-Kurn, Dan M.
and Gross, Christian
and Bloch, Immanuel},
title={A subradiant optical mirror formed by a single structured atomic layer},
journal={Nature},
year={2020},
month={Jul},
day={01},
volume={583},
number={7816},
pages={369-374},
issn={1476-4687},
doi={10.1038/s41586-020-2463-x},
url={https://doi.org/10.1038/s41586-020-2463-x}
}

@article{Shahmoon17,
  title = {Cooperative Resonances in Light Scattering from Two-Dimensional Atomic Arrays},
  author = {Shahmoon, Ephraim and Wild, Dominik S. and Lukin, Mikhail D. and Yelin, Susanne F.},
  journal = {Phys. Rev. Lett.},
  volume = {118},
  issue = {11},
  pages = {113601},
  numpages = {6},
  year = {2017},
  publisher = {American Physical Society},
  doi = {10.1103/PhysRevLett.118.113601},
  url = {https://link.aps.org/doi/10.1103/PhysRevLett.118.113601}
}

@article{Shahmoon24,
  title = {Dissipative transfer of quantum correlations from light to atomic arrays},
  author = {Ben-Maimon, Roni and Solomons, Yakov and Shahmoon, Ephraim},
  journal = {Phys. Rev. A},
  volume = {110},
  issue = {3},
  pages = {033719},
  numpages = {10},
  year = {2024},
  month = {Sep},
  publisher = {American Physical Society},
  doi = {10.1103/PhysRevA.110.033719},
  url = {https://link.aps.org/doi/10.1103/PhysRevA.110.033719}
}

@misc{SteckRb01,
author ={Steck, Daniel A.},
title = {Rubidium 87 {D} line data},
url = {https://steck.us/alkalidata/rubidium87numbers.1.6.pdf},
year = {2001}
}

@article{Tarbutt18,
author = {M. R. Tarbutt},
title = {Laser cooling of molecules},
journal = {Contemporary Physics},
volume = {59},
number = {4},
pages = {356--376},
year = {2018},
publisher = {Taylor \& Francis},
doi = {10.1080/00107514.2018.1576338},


}

@article{Vaaranta26,
title = {Numerical implementation of the partial secular approximation and unified master equation in structured open quantum systems},
journal = {Computer Physics Communications},
volume = {320},
pages = {109948},
year = {2026},
issn = {0010-4655},
doi = {https://doi.org/10.1016/j.cpc.2025.109948},
url = {https://www.sciencedirect.com/science/article/pii/S0010465525004497},
author = {Antti Vaaranta and Marco Cattaneo},
}

@Article{Wu2024,
author={Wu, Qilong
and Zhang, Yuan
and Wu, Hao
and Su, Shi-Lei
and Liu, Kai-Kai
and Oxborrow, Mark
and Shan, Chong-Xin
and M{\o}lmer, Klaus},
title={Theoretical study of superradiant masing with solid-state spins at room temperature},
journal={Science China Physics, Mechanics {\&} Astronomy},
year={2024},
month={May},
day={14},
volume={67},
number={6},
pages={260314},
issn={1869-1927},
doi={10.1007/s11433-023-2347-0},
url={https://doi.org/10.1007/s11433-023-2347-0}
}

@article{Yanes-Thomas25,
  title = {Collective coupling of driven multilevel atoms and its effect on four-wave mixing},
  author = {Yanes-Thomas, P. and Guti\'errez-J\'auregui, R. and Barberis-Blostein, P. and Sahag\'un-S\'anchez, D. and J\'auregui, R. and Kunold, A.},
  journal = {Phys. Rev. Res.},
  volume = {7},
  issue = {1},
  pages = {013028},
  numpages = {15},
  year = {2025},
  month = {Jan},
  publisher = {American Physical Society},
  doi = {10.1103/PhysRevResearch.7.013028},
  url = {https://link.aps.org/doi/10.1103/PhysRevResearch.7.013028}
}

\end{document}